\documentclass{article}
\usepackage{ijcai26}

\usepackage{times}
\usepackage{soul}
\usepackage{url}
\usepackage[hidelinks]{hyperref}
\usepackage[utf8]{inputenc}
\usepackage[small]{caption}
\usepackage{graphicx}
\usepackage{amsmath}
\usepackage{amsthm}
\usepackage{amssymb}
\usepackage{booktabs}
\usepackage{algorithm}
\usepackage{algorithmic}
\usepackage[switch]{lineno}
\usepackage{amsmath}
\usepackage{amsthm}
\usepackage{booktabs}
\usepackage{algorithm}
\usepackage{algorithmic}
\usepackage{subfigure}
\usepackage{makecell}
\usepackage{multirow}
\usepackage{color}
\usepackage{enumerate}

\title{OneVoice: One Model, Triple Scenarios—Towards Unified Zero-shot Voice Conversion}

\author{
Zhichao Wang
\and
Tao Li\and
Wenshuo Ge\and
 Zihao Cui\and
 Shilei Zhang\footnotemark[1]\And
 Junlan Feng\footnotemark[1] \\
\affiliations
JIUTIAN Research, China Mobile, Beijing, China \\
\emails
\{wangzhichao, zhangshilei, fengjunlan\}@cmjt.chinamobile.com
}

\begin{document}

\maketitle

\renewcommand{\thefootnote}{\fnsymbol{footnote}}
\footnotetext[1]{Corresponding author}
\renewcommand{\thefootnote}{\arabic{footnote}}
\begin{abstract}
Recent progress of voice conversion~(VC) has achieved a new milestone in speaker cloning and linguistic preservation. But the field remains fragmented, relying on specialized models for linguistic-preserving, expressive, and singing scenarios. We propose OneVoice, a unified zero-shot framework capable of handling all three scenarios within a single model. OneVoice is built upon a continuous language model trained with VAE-free next-patch diffusion, ensuring high fidelity and efficient sequence modeling.
Its core design for unification lies in a Mixture-of-Experts (MoE) designed to explicitly model shared conversion knowledge and scenario-specific expressivity. Expert selection is coordinated by a dual-path routing mechanism, including shared expert isolation and scenario-aware domain expert assignment with global-local cues. For precise conditioning, scenario-specific prosodic features are fused into each layer via a gated mechanism, allowing adaptive usage of prosody information. Furthermore, to enable the core idea and alleviate the imbalanced issue (abundant speech vs. scarce singing), we adopt a two-stage progressive training that includes foundational pre-training and scenario enhancement with LoRA-based domain experts. Experiments show that OneVoice matches or surpasses specialized models across all three scenarios, while verifying flexible control over scenarios and offering a fast decoding version as few as 2 steps. Audio samples are available on demo page\footnote{\url{https://kerwinchao.github.io/OneVoice/}}.

\end{abstract}

\section{Introduction}

Voice Conversion (VC) aims to convert the speaker timbre of a source speech while preserving its core content, i.e., linguistic information and prosodic features. This technology exhibits broad application value in human-computer interaction, entertainment, and privacy protection.
In recent years, VC technology has made notable progress: it can handle arbitrary input speech~\cite{PPGSun2016PhoneticPF} and achieve zero-shot capability to any target speaker~\cite{LM-VC,uniaudio}, pushing the boundaries towards generalized voice conversion.
However, challenges remain in processing diverse real-world speech, of which the focus of `content' is shifting from semantic-driven neutral reading, to expressive and emotional dubbing, and further to melody-guided singing performance. This shift raises a fundamental challenge in current VC research: how to understand and convert the multi-scenario core information within a unified framework, encompassing linguistic content, para-linguistic expressivity, and singing melody. This unified perspective is a crucial step toward generalized VC, and this work focused on the \textit{unified zero-shot VC paradigm}.

To meet diverse demands, current VC research has formed fragmented task paradigms, developing specialized models for distinct scenarios. Most existing works~\cite{PPGSun2016PhoneticPF,seedvc} focus on the general \textbf{Linguistic-preserving VC~(LVC)}, decoupling speech into linguistic content and speaker timbre for high speaker similarity and linguistic integrity. Another line of research emphasizes \textbf{Expressive VC~(EVC)}~\cite{Expressive-VC}, prioritizing the capturing of source prosodic information that captures diverse speaking styles and emotions. In contrast, \textbf{Singing VC~(SVC)}~\cite{svcc2023} ensures the converted result adheres to a given melodic contour. This fragmentation stems from the lack of a unified perspective to define which information should be `preserved' or `ignored' during conversion. Although recent studies, such as \cite{vevo2}, have attempted to construct unified systems by selecting unified features to encompass cross-scenario information, the overlooking of intrinsic differences in the unified representation and model design may cause inter-scenario interference, which can lead to performance degradation.

To avoid such interference and achieve unification with a unified perspective, several core challenges must be jointly considered:
\begin{itemize}
\label{sec:intro}
    \item \textbf{Modeling inter-scenario commonalities and differences}: linguistic content and speaker timbre are shared across VC scenarios, while diverse expressions attached to the linguistic content form the scenario specificity. The framework needs to capture shared elements while distinctly handling scenario-specific attributes -- such as para-linguistic prosody in expressive speech versus precise melodic contours in singing, which follow vastly different acoustic distributions between speech and singing~\cite{f0distru}.
    \item \textbf{Learning from highly imbalanced data}: there is a huge disparity in the scale of publicly available speech and singing data: speech corpora often encompass hundreds of thousands of hours, whereas singing data typically amounts to only several hundred hours. The framework should overcome this imbalance to ensure robust performance.
    \item \textbf{Maintaining efficiency without sacrificing fidelity}: 
    VC inherently preserves utterance duration, resulting in long semantic and acoustic sequences that challenge the efficiency of LM-based modeling. Deploying VC in practical applications requires overcoming the inefficiency of long-sequence modeling while maintaining the high generation quality expected of VC systems
\end{itemize}

To address these challenges, we propose OneVoice, a unified VC framework built upon a continuous LM integrated with a Mixture-of-Experts (MoE) structure. OneVoice achieves zero‑shot capability across all three scenarios: LVC, EVC, and SVC.
To balance modeling efficiency with generation quality, OneVoice is realized as a continuous LM trained with next-patch diffusion objective without the need for additional VAE tokenization. For unification, the core design of OneVoice explicitly models `shared' and `specialized' knowledge via dynamic expert routing: shared experts capture fundamental conversion abilities, while domain experts enhance scenario-aware expressivity. This design is further enabled by three components:
1) dual-path routing mechanism: shared expert isolation ensures that common knowledge is retained by the shared expert, while scenario‑aware expert assignment integrates a global prior (speech/singing mode) and local prosody‑context cues to activate domain experts in a scenario‑aware manner; 2) scenario-specific prosodic conditioning: the usage of scenario-specific prosody (shallow ASR features for EVC, discrete F0 for SVC) is adaptively adjusted by gated prosody fusion at each LM layer; 3) two-stage progressive training paradigm: we first conduct foundational pre-training on basic conversion, followed by a scenario enhancement stage for EVC and SV. This paradigm, along with LoRA-based domain experts, also effectively mitigates the data‑scale disparity between speech and singing.

Extensive experiments show that OneVoice matches or surpasses task-specific models across all three scenarios, confirming its effectiveness. Notably, OneVoice can flexibly switch conversion modes by disabling domain experts in the MoE or adjusting the global prior—either automatically or manually—to suit different requirements. When replacing the proposed flow-matching version of OneVoice with MeanFlow objective~\shortcite{meanflow}, OneVoice can also maintain high quality while achieving faster decoding~(e.g.,  RTF of 0.3 on an A100 GPU in 2 steps) without engineering optimizations. These results demonstrate the practical potential of OneVoice.

\section{Related Works}

\noindent\textbf{Voice Conversion.}
VC aims to alter the speaker timbre while preserving the essential information of the source speech. The definition of this `essential information' has continually expanded with research progress. Most prior work focuses on Linguistic-preserving VC~(LVC), which aims to perfectly disentangle speaker timbre from linguistic content to achieve high speaker similarity and intelligibility. The introduction of ASR-based~\cite{PPGSun2016PhoneticPF} and SSL-based~\cite{hsu2021hubert} linguistic features has enabled VC models to handle arbitrary input speech. Concurrently, by using a short clip of the target speaker as a prompt, both diffusion-based~\cite{seedvc,refvc} and LM-based~\cite{uniaudio,LM-VC} VC models can capture fine-grained speaker timbre, yielding impressive zero-shot results
To produce more expressive speech, Expressive VC~(EVC) has emerged, transferring para‑linguistic information from the source through additional conditioning signals—ranging from global~\cite{transferring} to fine‑grained~\cite{refvc} representations. Techniques such as information bottleneck~\cite{takinvc}, adversarial training~\cite{naturalspeech3}, or perturbation~\cite{nansy} are often employed to balance prosody modeling and speaker cloning, preventing timbre leakage from source speech in EVC. 
Differing from the above, Singing VC~(SVC)~\cite{svcc2023} focuses on singing voices, emphasizing adherence to the melodic contour of a song. It typically uses the fundamental frequency (F0) as the core conditional input. Methods like multi-scale F0 modeling~\cite{svc2023nwpu}, pre-trained pitch extractors~\cite{seedvc}, and data augmentation~\cite{YingMusic-SVC} are commonly used to enhance pitch modeling accuracy. Notably, the scale of publicly available singing data is far smaller than that of speech data, which limits the potential for zero-shot capability. As we can see, VC research is shifting from modeling solely linguistic information to handling multiple core aspects (linguistic content, para-linguistic prosody, and melodic prosody). This shifting creates a natural demand for a unified modeling framework. In this work, OneVoice aims to model both the inherent commonalities and the differences among these three scenarios within a single model.

\noindent\textbf{Mixture of Experts.}
The Mixture of Experts (MoE) architecture~\cite{smoe} has attracted much attention for its ability to efficiently scale model capacity by dynamically activating subsets of parameters. Its core is a routing network that selects the most suitable expert(s) for each input. Within an MoE layer, multiple independent experts are coordinated by a gating network (router). This router typically applies a trainable matrix to produce a probability distribution over the experts, normalized via a softmax function, and then applies strategies such as static top‑k selection or dynamic fusion to route tokens to selected experts. In the speech field, the MoE mechanism has been successfully applied to handle diverse attributes (e.g., dialect~\cite{HDMolE}, language\cite{languagemoe}, and cross‑modal interaction~\cite{moetts}), in both recognition and generation tasks. Various routing strategies, including unsupervised, task‑aware, and hierarchical routing, have been developed to direct inputs with different properties to dedicated experts. Inspired by \cite{deepseekmoe}, OneVoice adopts an MoE design that explicitly models `shared' and `specialized' knowledge with shared and domain experts, while a designed dual-path routing is introduced to enforce isolation of common knowledge while maintaining scenario awareness during expert selection. Furthermore, OneVoice follows a progressive training process that learns from commonality to scenario‑specific differences.

\noindent\textbf{Continuous Language Modeling.}
Language Models (LMs) have demonstrated powerful sequence modeling and zero-shot generalization capabilities in a range of generation tasks. Pioneering work~\cite{valle} has successfully applied the LM paradigm to audio generation by first discretizing audio into tokens using neural codecs~\cite{soundstream}, opening a new path toward high‑quality, zero‑shot speech synthesis. However, the long-sequence nature of audio hinders the efficiency of LM generation and cross-modal interaction, while codecs~\cite{moshi,llasa} face a trade-off between fidelity and compression.
Diffusion models, operating directly on compressed continuous features, excel at generating high-fidelity signals in domains like images and audio. Recent efforts~\cite{transfusion,msntd} combine the strengths of both: LM for sequence modeling and in-context learning, and diffusion for high-fidelity continuous generation, forming a `next-token diffusion' paradigm on continuous LM. Following this direction, models such as VibeVoice~\shortcite{vibevoice}, and DiTAR~\shortcite{ditar} further illustrate the potential of next-token diffusion in high-fidelity speech generation. Inspired by these advances, OneVoice adopts a continuous LM with a diffusion head as its backbone and further employs a patch-based modeling strategy~\cite{ditar} to compress sequence length, resulting in \textit{next-patch diffusion} paradigm, enabling both efficient and high-fidelity conversion.

\begin{figure}[htbp]
\vspace{-5pt}
 \centering
    \includegraphics[width=1\linewidth]{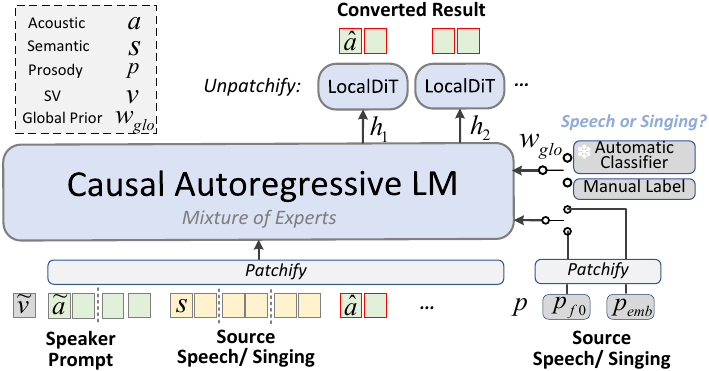}
 \caption{The overall architecture for OneVoice.}
	\label{fig:onevoice}
 \vspace{-5pt}
\end{figure}

\section{OneVoice}

As depicted in Fig.~\ref{fig:onevoice}, OneVoice is built upon a causal LM with a built-in local DiT~\shortcite{dit} as its diffusion head. In this framework, speech and singing are represented as semantic feature $\mathbf{s}\in\mathbb{R}^{T\times D_s}$ and acoustic feature $\mathbf{a}\in\mathbb{R}^{T\times D_a}$ by an ASR encoder and mel-spectrogram estimation. Here, $T$ denotes the sequence length while $D_s$ and $D_a$ represent the corresponding feature dimensions. The prosody features $\mathbf{p} \in \{\mathbf{p}_{f0},\mathbf{p}_{emb}\}$ are  extracted from the source utterance. 
Given a speaker prompt $ \{\mathbf{\tilde{v}},\mathbf{\tilde{a}}\}$ from the target speaker, OneVoice can assign MoE experts based on pre-trained automatic classifier or manual label~(singing or speech), converting the input semantic features $\mathbf{s}$ and prosody $\mathbf{p}$ into mel results $\mathbf{a}$ for EVC or SVC mode, while the discarding of the domain experts and prosody can activate LVC mode. During the conversion, the LM process at the patch level, and the LocalDiT head subsequently performs non-autoregressive un-patchify to reconstruct the final mel-spectrogram.

\begin{figure}[htb]	
\vspace{-10pt}	
 \centering
    \includegraphics[width=0.9\linewidth]{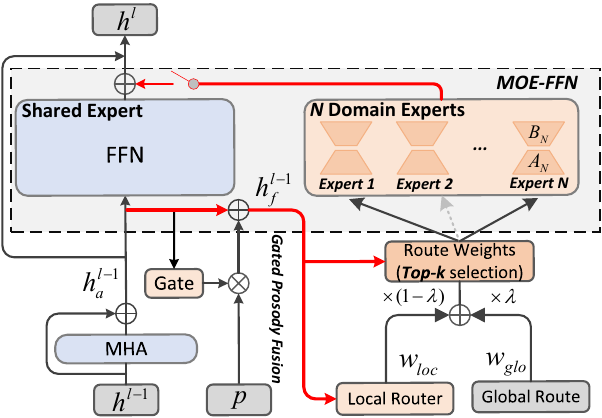}
 \caption{The details of LM block.}
	\label{fig:block}
    \vspace{-10pt}	
\end{figure}

\subsection{Conditional MoE Architecture}
\label{sec:moe}

As outlined in Section~\ref{sec:intro}, diverse expressions—such as EVC and SVC —can be viewed as deriving from the fundamental LVC by augmenting linguistic content with either para-linguistic prosody or melodic contours. This perspective guides the core design of OneVoice. To jointly model shared and scenario‑specific knowledge, we employ a language model integrated with a conditional MoE as the backbone. While MoE can dynamically activate different parameter subsets for inputs from different scenarios, traditional MoE suffers from knowledge redundancy among experts~\cite{deepseekmoe}, which contradicts our goal of separating shared and specialized knowledge. Moreover, biased router assignments may lead to ambiguous domain-expert correspondence~\cite{HDMolE}, causing inter-scenario interference. To enable our goal, we introduce explicit shared-domain expert specialization together with a dual‑path routing mechanism  (Shared Expert Isolation \& Scenario-aware Expert Assignment), as illustrated in Fig.~\ref{fig:block}.

\textbf{Shared-domain Expert Specialization.} Following the idea of separating common and specialized knowledge, each MoE layer contains two distinct types of experts: one shared expert $f^{s}$ and $N$ domain experts $\{f^d_i\}_{i=1}^{N}$. 
Given the $h_a^l$ as the common input to the MoE layer, the shared expert is responsible for capturing foundational capabilities common to all VC scenarios, such as content preservation and speaker cloning. In parallel, the domain experts, guided by scenario-specific knowledge $h_f^l$, enhance the modeling of attributes specific to EVC or SVC.

\textbf{Dual-path Routing Mechanism.} In OneVoice, two routing strategies, shared expert isolation and scenario-aware expert assignment, are designed to steer the specialization of shared and domain experts, respectively. For \textit{shared expert isolation}, we assign one fixed expert as the shared expert. Its role is to capture common knowledge across all inputs, independently of the router’s decisions. For domain experts, to resolve ambiguous domain-expert correspondence, we employ a \textit{scenario-aware expert assignment} that combines: 1) global routing: domain experts are grouped into two parts and assigned initial weights according to a patch-level global prior (speech or singing mode), which can be provided manually or predicted automatically. This yields a global routing weight $w_{glo} \in \mathcal{R}^{T^\prime \times N}$ with patch-level sequence length $T^\prime$, encouraging experts to focus on prior scenario knowledge. Please note that the global prior is a continuous probability over the two scenarios. For example, if the current patch has a singing probability of 0.8 (0.2 for speech), each of the two experts in the singing group will receive a global weight of $0.4=0.8/2$;
2) local routing: to enhance sensitivity to dynamic input, we fuse the current hidden state $h_a^l$ with the scenario-specific prosodic condition $p$ (See Section~\ref{sec:prosody}). A local router then computes a local routing weight $w_{loc}$ based on the fused hidden state $h_f^l$. The final routing weight 
$w$ is obtained by a linear combination: $w=\lambda w_{glo}+(1-\lambda)w_{loc}$ , where $\lambda$ is a balancing factor. Finally, top-K experts with the highest weights are selected to process the current input. 

With the above design, the output $h^l$ of MOE-FFN block can be expressed as:
\vspace{-10pt}
\begin{equation}
\centering
    h^{l}={f}^{s}(h^{l}_{a})+\sum_{i=1}^{N}(g_{i}{f}_{i}^{d}(h_{f}))+h^{l}_{a},
\end{equation}
\begin{equation}
\centering
    g_{i}=\left\{\begin{matrix}
  w_{i}, &w_{i}\in TopK(\{w_{j}|1\le j \le N\},K)\\
  0,   &otherwise
\end{matrix}\right. ,
\end{equation}
\begin{equation}
\centering
    w_{i}=(1-\lambda)w_{loc,i}+\lambda w_{glo,i},
\end{equation}
\begin{equation}
\centering
    w_{loc,i} = {Softmax}_i(\mathbf{W}h^{l}_{f}),
\end{equation}
where $g_i$ denotes the gate value for the $i$-th expert and $\mathbf{W}$ represents the trainable weights of the local router. This design enables dynamic, scenario-aware expert selection by jointly considering static global prior and dynamic local cues. Additionally, we apply a load-balancing loss $\mathcal{L}_{balance}$~\cite{switch} to the local router weights for a more balanced distribution for further expert selections.

\subsection{Scenario-Specific Prosodic Conditioning}
\label{sec:prosody}

Building upon the foundation of LVC, EVC and SVC further emphasize the modeling of para-linguistic expressivity and melodic contours, respectively. As mentioned in the introduction, the inherent acoustic differences between speech and singing motivate our design of separate, scenario-specific prosodic conditioning, which enhances modeling accuracy and reduces inter‑scenario interference.

\textbf{Scenario-Specific Prosody.} For EVC, we leverage shallow-layer features $p_{emb}$, which retain rich prosody information~\cite{YZ}, extracted from a pre-trained ASR encoder as the prosodic condition. To prevent leakage of speaker identity, perturbation~\cite{nansy} and bottleneck techniques are employed during training. Additionally, a masking strategy~\cite{hsu2021hubert} is applied to $p_{emb}$ to ensure the model relies solely on prosody rather than linguistic information in $p_{emb}$. In terms of SVC, following \cite{seedvc}, we quantize the F0 extracted by a robust pitch estimator (RMVPE) into a 256-bin discrete sequence. During inference, F0 sequence is adaptively shifted according to the target speaker's pitch statistics~\cite{YingMusic-SVC} and the mean pitch of the target singer. The choice of which prosodic feature to use is determined by the patch‑level global prior.

\textbf{Gated Prosody Fusion.} To effectively integrate prosodic conditions, we employ a gated fusion mechanism in each LM block. The key idea is to allow the model to adaptively regulate the influence of scenario-specific prosody based on the current hidden state, instead of fixed addition or concatenation. As illustrated in Fig.~\ref{fig:block}, for the $l$-th LM block, the gating network $\mathcal{G}^{(l)}$ first computes a gating value $\mathbf{g}^l \in \mathbb{R}^{T^\prime}$ from the hidden state $\mathbf{h}_a^l$ output by the attention module:
\begin{equation}
\mathbf{g}^l = \sigma(\text{Linear}(\mathbf{h}_a^l)),
\end{equation}
where \(\sigma\) is the sigmoid function. The patchified prosodic sequence \(\mathbf{P}^\prime\) is then modulated by this gate and fused with the hidden state:
\begin{equation}
    \mathbf{h}_f^l = \mathbf{h}_a^l + \mathbf{g}^l \odot \mathbf{P}^\prime.
\end{equation}

This gating design enables the model to utilize prosodic information with varying behaviors at different layers. Furthermore, the prosody-augmented hidden state \(\mathbf{h}_f^l\) is also fed to the local router (Section~\ref{sec:moe}), providing finer-grained scenario cues to guide expert selection.

\subsection{VAE-free Next Patch Diffusion}
Inspired by the success of patch-based hierarchical LM~\cite{megabyte} in long-context tasks, recent work~\cite{ditar} further extends this paradigm to speech generation by integrating continuous diffusion modeling, leveraging its strengths in long‑range coherence and fine‑grained acoustic detail. Following DiTAR~\shortcite{ditar}, we build OneVoice under a \textit{`next-patch diffusion'} paradigm, consisting of a patchify layer, a core LM, and a LocalDiT head, for efficient and high-fidelity generation. The LM operates on patch‑level sequences and produces a hidden state $h_t$ at each step, which is then passed to the LocalDiT head for un‑patchify. Notably, rather than relying on a VAE to produce compressed continuous latents,
we directly use the mel‑spectrogram as the acoustic feature $\mathbf{a}$ for LocalDiT prediction. Mel spectrogram preserves full acoustic interpretability and aligns better with human auditory perception. The \textit{VAE-free} design also simplifies the generation pipeline (LM$+$vocoder) without additional VAE training. With patch‑based compression, the LM can operate at 10Hz and even lower frame rates, significantly reducing sequence length.

\textbf{Patchify.} Before inputting to the LM, the layer of patchify firstly converts the original sequence, consisting of semantic, acoustic, and prosody features$\{\mathbf{s},\mathbf{a},\mathbf{p}\}$, into a patch level sequence $\{\mathbf{a}^\prime,\mathbf{s}^\prime,\mathbf{p}^\prime\}$ with distinct patch ratio $r$. For simplicity, all features are patchified into the same length of $T^\prime$ after linear embedding into continuous space. For example, acoustic feature $\mathbf{a}$~($T \times D_a$) is linearly transformed and reshaped into a sequence of size $\frac{T}{r} \times r D^\prime $, and then projected back to dimension $D^\prime$ with length $T^\prime = \frac{T}{r}$. 

\begin{figure}[htb]	
 \centering
    \includegraphics[width=0.6\linewidth]{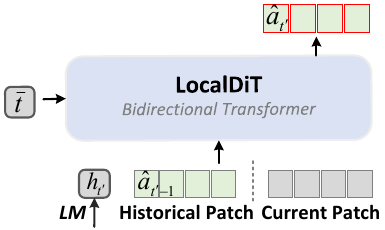}
 \caption{The LocalDiT block.}
	\label{fig:localdit}
 \vspace{-5pt}	   
\end{figure}

\textbf{LocalDiT with Flow Matching.} As shown in Fig.~\ref{fig:localdit}, following the design of DiTAR~\shortcite{ditar}, we employ a bidirectional local transformer as the diffusion head, called LocalDiT. Given the hidden state $h_{t^\prime}$ from LM, LocalDiT iteratively produces the mel spectrogram within $t^\prime$-th patch over diffusion time $\bar{t}$, where the last historical patch is also employed as a prefix input for better generation quality. This modeling process can be formulated as: 
$p_{\theta}(\mathbf{a}_{r\cdot {t^\prime}+1:r\cdot {t^\prime}+r}|\bar{t} ,h_{t^\prime},\mathbf{a}_{r \cdot ({t^\prime}-1)+1:r \cdot ({t^\prime}-1)+r})$.
Let $f_{\theta}(\cdot)$ denote the LocalDiT network. OneVoice is optimized with a conditional flow-matching objective within Optimal Transport (OT) displacement map~\cite{flowmatching}:
\begin{equation}
    \mathcal{L}_{FM} = \mathbb{E}_{\bar{t},a_1,\epsilon} \left [ \left \|  v_\theta(a_{\bar{t} },\bar{t} )- v(a_1,\epsilon)\right \|^2_2 \right ],
\end{equation}
where
\begin{equation}
    v_\theta(a_{\bar{t}},\bar{t} ) = f_{\theta}{(a_{\bar{t}},\bar{t},c)}, \quad v(a_1,\epsilon)= a_1-\epsilon,
\end{equation}
\begin{equation}
    a_{\bar{t}} = (1-\bar{t})\epsilon + \bar{t}a_1.
\end{equation}
Here $\bar{t} \in [0,1]$, $\epsilon \sim \mathcal{N}(0,I)$ is standard Gaussian noise, and $a_1 \sim q(a_1)$ is the data samples. To emphasize the conditional signal, the LM guidance $h_{t^\prime}$ is randomly masked in LocDiT and we enable classifier-free guidance (CFG) during inference: $v_{cfg}=(1+\gamma) f_{\theta}(a_{\bar{t}},\bar{t},c) - \gamma f_{\theta}(a_{\bar{t}},\bar{t},\emptyset)$. Notably, we also explore replacing flow matching with \textit{MeanFlow}~\shortcite{meanflow} in OneVoice for faster sampling in experiments.

\subsection{Two-stage Progressive Training Paradigm}
\label{sec:training}

To equip OneVoice with both general conversion ability and specialized expressivity from imbalanced data, we employ a two-stage training strategy that progressively enhances the model while preserving foundational knowledge.
1) Foundational pre-training: The entire model with shared expert (blue modules in Fig.~\ref{fig:block}) is first trained on a large-scale corpus for the core LVC task, establishing a robust base for speaker cloning and content preservation; 2) Scenario-aware Enhancement: the pre‑trained model is then continually trained to progressively acquire EVC and SVC abilities. We employ a parameter‑grouped layer‑wise learning‑rate schedule: newly introduced domain‑expert parameters (orange modules in Fig.~\ref{fig:block}) are updated with a higher learning rate to learn scenario‑specific knowledge, while the previously learned parameters (blue modules) are fine‑tuned with a much lower rate, allowing them to adaptively collaborate with the domain experts in complex scenarios. Besides, in the 2nd stage, to mitigate the data imbalance between abundant speech and scarce singing samples, each training batch maintains an equal sampling ratio of speech and singing utterances. Moreover, during this stage, domain‑expert usage is randomly dropped with a probability of 0.3 to keep the foundational LVC knowledge. The global prior (speech/singing mode) is provided via manual labels with label smoothing; for robustness, we also switch the label (singing$\longleftrightarrow$speech) with a probability of 0.1. In practice, all domain experts are implemented using LoRA~\shortcite{lora} for efficient learning. With the paradigm, OneVoice can alleviate imbalanced issues and achieve natural shared-specialized innovation for LVC, EVC, and SVC.

\section{Experiments}

\subsection{Experimental Setup}
\textbf{Corpus.} Training data of OneVoice comprises two main categories: 1) Speech corpus: a open-sourced Emilia~\shortcite{emilia} comprising \textbf{100,000} hours is used; 2) Singing corpus: we collected about \textbf{400} hours singing data from open-sourced projects, including PopBuTFy~\shortcite{PopButFy}, ACEsinger~\shortcite{ACEsinger}, M4Singer~\shortcite{M4Singer}, OpenCpop~\shortcite{opencpop}, GTSinger~\shortcite{GTsinger}, OpenSinger~\shortcite{Opensinger}, and NUS-48E~\shortcite{nus48e}. For evaluation, we construct three testsets (LVC, EVC, and SVC), each consisting of 400 source–target utterance pairs. The source utterances cover a variety of expression forms, such as neutral reading, emotional speech, movie dubbing, and singing. The target speaker/singer utterances come from seedtts-eval\footnote{https://github.com/BytedanceSpeech/seed-tts-eval} and a self-collected dataset.

\noindent\textbf{Implement Details.}
\label{sec:configuration}
For acoustic representation, a 24kHz waveform is processed into an 80‑dim mel‑spectrogram at a frame rate of 50Hz. Mel‑to‑waveform reconstruction is performed using BigVGAN~\shortcite{bigvgan} from kimi-audio\footnote{\url{https://huggingface.co/moonshotai/Kimi-Audio-7B}}. For prosody and semantic features $\{p_{emb},s\}$, we extracted 50Hz bottleneck features of the 6th and last encoder layer from an ASR\footnote{\url{https://github.com/wenet-e2e/wenet/tree/main/examples/wenetspeech/s0}\label{model:nsasr-wenetspeech}} trained on WenetSpeech. Speaker embedding $v$ is extracted from CAM++\footnote{\url{https://github.com/modelscope/3D-Speaker/tree/main/egs/3dspeaker/sv-cam}}.
OneVoice employs six Transformer blocks with 8 attention heads and a hidden size of 1024 as its core LM. Each MoE‑FFN layer consists of 1 shared expert (the original FFN) and 6 domain experts, with $K=2$ domain experts selected during routing. The LoRA-based experts use 32-rank and $\alpha=1$. The balancing factor $\lambda$ in the local router is $0.5$. The LocalDiT head contains 4 DiT blocks, each with 8 heads and a hidden dimension of 1024. The patchify layer is implemented via a linear projection with a patch ratio $r=5$ for $\{a,s,p\}$, resulting in 10Hz sequence.
For prosodic modeling with $p_{emb}$, we apply a masking strategy with a mask ratio ranging from $0.01$ to $0.02$ and a span length of 10. A bottleneck layer compresses the 512-dim $p_{emb}$ into a 32-dim feature. 
 During training, the LM conditions for LocalDiT are dropped with a probability of 0.1, while inference uses 8 diffusion timesteps with the 0.5 cfg ratio. The max training length is set to 30s. 8 A100 GPUs are used to pre-train the blue part (Fig~\ref{fig:block}) for 600k steps with $1 \times 10^{-4}$ learning rate. The subsequent enhancement is performed on 4 GPUs for 200k steps, where the newly added orange modules are trained with a learning rate of $1 \times 10^{-4}$ and the pre‑trained blue modules are fine‑tuned with a learning rate of  $1 \times 10^{-5}$.

\noindent\textbf{Evaluation Metrics.}
Content Enjoyment (A-CE) from Meta Audiobox Aesthetics\footnote{\url{https://github.com/facebookresearch/audiobox-aesthetics}} automatically measures subjective audio quality. Character error rate (CER) measured by FireRedASR\footnote{\url{https://github.com/FireRedTeam/FireRedASR}} indicates the speech intelligibility. Speaker similarity (SSIM) is calculated by an open-sourced SV model\footnote{\url{https://www.modelscope.cn/models/iic/speech_eres2net_large_sv_zh-cn_3dspeaker_16k}}. Besides, SER-based cosine similarity\footnote{\url{https://www.modelscope.cn/models/iic/emotion2vec_base/}} (PSIM) and F0 Correlation Coefficient (F0$_{cor}$) are used to measure the prosody consistency for EVC. Comparative mean opinion score subjectively measures the speaker similarity~(CMOS$_s$) and prosody consistency~(CMOS$_p$) between OneVoice and comparison systems, in which CMOS $>0$ means OneVoice is better and vice versa.

\subsection{Experiments Results}
\label{sec:experiments}

We select recent representative VC systems across three distinct scenarios for comparison: SeedVC~\shortcite{seedvc} and Metis-VC~\shortcite{Metis} for LVC, Vevo~\shortcite{vevo} and REF-VC~\shortcite{refvc} for EVC, and SeedVC-Sing and YINGSVC~\shortcite{YingMusic-SVC} for SVC. Notably, the results of REFVC are obtained from the author, while the officially open-sourced codes and models of others can be found. We implement the proposed system OneVoice integration flow-matching objective, while the 2-step MeanFlow (MF) version also involves the evaluation.

\begin{table}[htp]
\centering
\footnotesize
\setlength{\tabcolsep}{0.7mm}
\renewcommand\arraystretch{1.2}
\begin{tabular}{lcccc}
\hline
Method   &A-CE$\uparrow$ & CER $\downarrow$ & CMOS$_s$ $\downarrow$  & SSIM $\uparrow$  \\ \hline
 \multicolumn{3}{l}{\textit{\textcolor[RGB]{57,57,57}{LVC Models}}}       &   &     \\ 
  SeedVC  & 4.91    &1.27  & \textbf{-0.17} & \textbf{0.733}      \\
 Metis-VC&  4.90    & 6.49  & 0.47    & 0.678         \\\hline
 \multicolumn{3}{l}{\textit{\textcolor[RGB]{57,57,57}{In LVC Mode}}}          &  &   \\ 
 OneVoice   &  \textbf{5.11}    & \textbf{0.88}  & 0 & 0.729   \\
  $\quad  +$  \textit{MF}  & 4.96   & 1.32  &0.10 & 0.711    \\
 \hline
\end{tabular}
\caption{Zero-shot LVC performance}
\label{exp:LVC}
\end{table}

\textbf{LVC Evaluation.} Table~\ref{exp:LVC} presents the results in LVC scenario. Compared to specialized LVC models, OneVoice in LVC mode achieves better results in terms of quality and intelligibility, while attaining speaker similarity close to SeedVC, indicating strong zero‑shot LVC performance. 
The MeanFlow variant of OneVoice shows a slight degradation relative to the flow‑matching version, yet it still delivers comparable performance while enabling faster decoding.

\begin{table}[htp]
\centering
\footnotesize
\setlength{\tabcolsep}{0.1mm}
\renewcommand\arraystretch{1.2}
\begin{tabular}{lccccccc}
\hline
Method   &A-CE$\uparrow$ & CER $\downarrow$ & CMOS$_s$ $\downarrow$ & SSIM $\uparrow$  & CMOS$_p$$\downarrow$   & F0$_{cor}$  $\uparrow$ & PSIM $\uparrow$  \\ \hline
 \multicolumn{3}{l}{\textit{\textcolor[RGB]{57,57,57}{In LVC Mode}}}          &  &   \\ 
 OneVoice&   5.42  & 2.23  & - &  0.725   & - & 0.708 & 0.486       \\ \hline
  \multicolumn{3}{l}{\textit{\textcolor[RGB]{57,57,57}{EVC Models}}}          &  &  &   & \\ 
   Vevo  &  5.25   & 4.07 &  0.32  & 0.650 & 0.57  & 0.750  & 0.589   \\
 REF-VC  &  5.43  & 1.87  & 0.50 &  0.639  & \textbf{-0.39}  & 0.796  &  \textbf{0.625} \\
 \hline
 \multicolumn{3}{l}{\textit{\textcolor[RGB]{57,57,57}{In EVC Mode}}}          &  &   \\ 
 OneVoice   &  \textbf{5.48}    & \textbf{1.38}  &  \textbf{0} &  \textbf{0.674}  & 0 & \textbf{0.800}   & 0.587 \\
  $\quad  +$  \textit{MF}  & 5.21  & 1.72  & 0.12  &  0.671  & 0.42 & 0.746   & 0.527 \\
 \hline
\end{tabular}
\caption{Zero-shot EVC performance}
\label{exp:EVC}
\end{table}

\begin{table*}[htp]
\centering
\footnotesize
\setlength{\tabcolsep}{2mm}
\renewcommand\arraystretch{1.1}
\begin{tabular}{lcccc|ccc|cc}
\hline
\multicolumn{1}{l}{\multirow{2}{*}{Method}} & \multicolumn{4}{c}{EVC} & \multicolumn{3}{c}{SVC} & \multicolumn{2}{c}{LVC} \\ \cline{2-10}
  & \multicolumn{1}{c}{CER} $\uparrow$  & SSIM $\uparrow$   &  F0$_{cor}$ $\uparrow$   & PSIM  $\uparrow$ & CER $\uparrow$  & SSIM $\uparrow$   &  F0$_{cor}$ $\uparrow$     & CER $\uparrow$  & SSIM $\uparrow$ \\ \hline
OneVoice & 1.38  & 0.674  & 0.800 & 0.587  & 3.92 &  0.738   & 0.898 &  0.88 & 0.729 \\ \hline
 \multicolumn{10}{c}{\textit{\textcolor[RGB]{57,57,57}{Prosodic Conditioning}}}     \\   \hline
  $\quad$ \textit{w/o} Gate Unit & 4.62  & 0.648  & 0.759 & 0.571 & 5.74  &  0.694 & 0.815 & - & -  \\
 $\quad$ \textit{w/o} Multi-layer Fusion & 3.32  & 0.660  & 0.728 & 0.536 & 6.98  &  0.700 & 0.792 & - & -   \\ \hline
 \multicolumn{10}{c}{\textit{\textcolor[RGB]{57,57,57}{Domain Expert Assignment}}}        \\ \hline 
$\quad$ \textit{w/o} Global Routing & 1.35  & 0.659  & 0.786 & 0.564 & 4.20  &  0.694 & 0.847 & - & -   \\
 $\quad$ \textit{w/o} Local Routing & 1.46  & 0.660  & 0.763 & 0.535 & 4.01  &  0.695 & 0.830 & - & -  \\ \hline
   \multicolumn{10}{c}{\textit{\textcolor[RGB]{57,57,57}{LoRA-involved Training}}} \\ \hline
  $\quad$ \textit{w/} Frozen Backbone & 1.69 & 0.690  & 0.771 & 0.510 & 4.17  &  0.710 & 0.823& 1.05 & 0.738  \\
   $\quad$ \textit{w/} 128-rank LoRA & 1.32  & 0.687  & 0.795 & 0.567 & 4.17  &  0.719 & 0.835 & - & -  \\ \hline
 \multicolumn{10}{c}{\textit{\textcolor[RGB]{57,57,57}{Patch Size}}}        \\ \hline 
  $\quad$ \textit{w/} Patch~$r=10$ & 3.21  & 0.674  & 0.688 & 0.500 & 4.70  &  0.710 & 0.748 & 1.42 & 0.694  \\
   \hline
   
\end{tabular}
\caption{Component Analysis}
\label{exp:ablation}
\end{table*}

\textbf{EVC Evaluation.} As shown in Table~\ref{exp:EVC}, OneVoice obtains superior results in most subjective and objective metrics.
Notably, REFVC gets higher prosody consistency (PSIM and CMOS$_p$), but yields lower speaker similarity~(SSIM and CMOS$_s$), compared to OneVoice. This can be attributed to the speaker leakage problem encountered in EVC when facing highly expressive source utterances. It indicates that OneVoice can get a better balance between voice cloning and prosody transfer. In the 2nd row, we also include OneVoice's results in the LVC‑mode on the EVC test set, which further clarifies the distinct objectives and behavior of LVC versus EVC models.
In LVC, without additional source prosody, the model tends to follow the timbre and speaking style of the speaker prompt, thus achieving a higher speaker similarity. In contrast, EVC mode must maintain speaker similarity while transferring the prosody of the source speech, which may differ significantly from that of the speaker prompt. These differences observed a trade‑off between speaker similarity and prosodic consistency. 
These distinct characteristics address different practical needs, and OneVoice’s switchable design provides a flexible way to meet them. And interestingly, benefiting from the large-scale dataset, we found that the LVC-mode of OneVoice can also convert non-verbal sound (e.g., laughter, coughs), which challenges the common assumption that ASR‑driven features trained with text‑based losses discard such non‑linguistic information.

\begin{table}[htp]

\centering
\footnotesize
\setlength{\tabcolsep}{0.4mm}
\renewcommand\arraystretch{1.2}
\begin{tabular}{lcccccc}
\hline
  Method  &A-CE$\uparrow$   & CER  $\downarrow$ & CMOS$_s$$\downarrow$ & SSIM $\uparrow$   & CMOS$_p$$\downarrow$ &  F0$_{cor}$ $\uparrow$  \\ \hline
 \multicolumn{2}{l}{\textit{\textcolor[RGB]{57,57,57}{SVC Models}}}    &   &   &  &   &     \\  
 YINGSVC &  5.96  & 4.26   & 0.64 & 0.705  &      0.36     & \textbf{0.911} \\ 
  SeedVC-Sing &  5.79  &  4.30 & 0.18 &  \textbf{0.745}  & 0.51 & 0.895    \\
 \hline
 \multicolumn{4}{l}{\textit{\textcolor[RGB]{57,57,57}{In SVC Mode}}}      &   & &        \\ 
 OneVoice    &  \textbf{6.15}   & \textbf{3.92} & \textbf{0}   & 0.738 &    \textbf{0} & 0.898    \\
 $\quad  +$  \textit{MF}  & 5.92  & 4.21 & 0.28 & 0.713   &   0.66 & 0.883  \\  \hline
 
\end{tabular}
\caption{Zero-shot SVC performance}
\label{exp:SVC}
\vspace{-5pt}
\end{table}

\textbf{SVC Evaluation.} As illuminated in Table~\ref{exp:SVC}, both objective and subjective results show that OneVoice delivers better singing expression and conversion quality while maintaining high speaker similarity. Besides, the MF version can also achieve comparable results with the comparison systems in the SVC task. 
In practice, we found that the strategy of transferring F0 from source speaker to target speaker in inference plays an important role in SVC model, directly affecting the speaker similarity and singing quality. In the SVC mode of OneVoice, we employ an adaptive pitch shift from YINGSVC~\shortcite{YingMusic-SVC} for general target speakers and a mean-value shift of F0 for target singers. 

Beyond conversion performance, the proposed 8-step flow-matching and 2-step MeanFlow of OneVoice can obtain RTF of 0.68 and 0.37 on a single A100 GPU, respectively. These results confirm that OneVoice matches or surpasses specialized models across three scenarios. And the multi-scenario support of OneVoice shows the practical potential to meet diverse applications.

\subsection{Discussion: Component Analysis}
\textbf{Prosodic Conditioning.} As shown in Table~\ref{exp:ablation}, removing the gating units from each layer degrades both CER and SSIM in EVC and SVC. This decline occurs because directly injecting prosodic signals into every layer disturbs the linguistic and timbre information encoded in the hidden states, thereby harming conversion quality.
When we simply keep one prosodic fusion layer in OneVoice, the model can't capture enough prosody information during conversion. This insufficient prosody modeling also limits the pronunciation accuracy in high-expressive conversion, especially for the SVC scenario. These observations indicate the important role of signal conditioning in the conversion framework.

\textbf{Domain Expert Assignment.}
In OneVoice, we use the domain expert assignment with global-local routing for both scenario awareness and expert utilization. When the global prior is removed, objective results for both EVC and SVC can be observed a degradation, indicating the effectiveness of introducing scenario prior. In contrast, discarding local routing, the same as using fixed scenario-specific parameters, limits the model capacity and dynamic processing, resulting in worse performance of prosody transfer. 

\textbf{LoRA-involved Training.}
As described in Section~\ref{sec:training}, OneVoice mainly obtains SVC and EVC abilities in the 2nd training stage, given the imbalanced data. We verified the key training and LoRA settings in this LoRA-involved training process. In \textit{w/ Frozen Backbone}, we freeze the parameters learned during foundational pre-training and update only the newly introduced module (orange part in Fig.~\ref{fig:block}). Table~\ref{exp:ablation} shows that this variant well preserves the foundation ability, in terms of SSIM and CER, with fast training speed, but it limits the prosody modeling (F$_{cor}$ and PSIM) in SVC and EVC. This limitation arises because the frozen pre‑trained parameters are less adaptable to generating highly expressive or singing voices.
In contrast, OneVoice, which fine‑tunes all parameters with a lower learning rate for the pre‑trained components, retains strong LVC performance while achieving better prosody transfer. Besides, increasing the LoRA rank from 32 to 128 did not yield significant performance gains under the current dataset scale, suggesting that a 32-rank is sufficient for the current available data.

\textbf{Patch Size.}
In addition to training OneVoice on a patch ratio of 5, resulting in 10Hz sequences, we also implement a version with a more aggressive patch ratio of 10. The results are presented in Table 4, which shows a rapid performance degradation in terms of CER, F0$_{cor}$, and PSIM across all three scenarios. This patch ratio of 10 makes the LM of OneVoice operate at 5Hz sequences, which may already lose the information integrity within each patch and necessitate increased LocalDiT parameters and LM hidden size. A similar phenomenon is also reported in DiTAR~\shortcite{ditar}.

\section{Conclusions}

In this work, we propose OneVoice, a unified voice conversion framework capable of handling three distinct scenarios—linguistic-preserving, expressive, and singing conversion—within a single model. By leveraging a MoE architecture with explicit shared and domain expert specialization, dual-path routing, and scenario-specific prosodic conditioning, OneVoice effectively captures both common and scenario-specific knowledge. The two-stage progressive training paradigm, combined with VAE-free next-patch diffusion, enables efficient and high-fidelity generation while mitigating data imbalance between speech and singing domains. Extensive experiments demonstrate that OneVoice achieves performance comparable to or surpassing task-specific models across all three scenarios, while offering flexible mode switching and fast inference. Our work represents a step towards unified zero-shot voice conversion, paving the way for more adaptive and scalable conversion systems.

\textbf{Limitations and Future Work.}  
OneVoice still has several limitations. Although it performs competitively across the three scenarios, the conversion results still have some mismatch with subjective preferences, and future work could build preference data for further preference optimization. Additionally, the current designed non-streaming architecture restricts real-time streaming applications.
Our future work will continually use more training data to explore the model architecture and performance potential of OneVoice. And OneVoice will extend the streaming ability by interleaved sequence modeling.

\newpage
\appendix

\bibliographystyle{named}
\bibliography{ijcai26}

\end{document}